\documentclass[prl,twocolumn,showpacs,amsmath,amssymb,superscriptaddress]{revtex4}
\usepackage{pifont,amssymb,epsf,graphicx}
\usepackage{amssymb}
\usepackage{txfonts}
\usepackage{bbm}
\usepackage{graphicx}
\usepackage{appendix}
\usepackage{epsf}
\usepackage{epstopdf}
\usepackage{amsmath}
\usepackage[usenames]{color}

\begin{document}

\title{Superfluid Density in the $s_{\pm}$ Wave state of Clean Iron-based Superconductors}

\author{Huaixiang Huang}
\affiliation{Department of Physics, Shanghai
University, Shanghai 200444, China}
\affiliation{Texas Center for Superconductivity and Department of
Physics, University of Houston, Houston, Texas 77204, USA}
\author{ Yi Gao}
\affiliation{Department of Physics and Institute of Theoretical Physics, Nanjing Normal University, Nanjing, Jiangsu 210046, China}
\author{ Jian-Xin Zhu}
\affiliation{Theoretical Division and Center for Integrated Nanotechnologies, Los Alamos National Laboratory, Los
Alamos, New Mexico 87545, USA}
\author{ C. S. Ting}
\affiliation{Texas Center for Superconductivity and Department of
Physics, University of Houston, Houston, Texas 77204, USA}
\date{\today}
\begin{abstract}
Based on a phenomenological model and the Kubo formula, we
investigate the superfluid density $\rho_s(T)$ and then the
penetration depth $\lambda(T)$  of the iron-based superconductors in the coexistence region of the spin-density wave and superconductivity, and also in  the overdoped region.
Our calculations show a dramatic increase of $\lambda(0)$ with the
decrease of the doping concentration $x$ below $x=0.1$. This result is consistent with the experimental observations. At low
temperatures, $\rho_s(T)$ shows an exponential-law behavior, while at
higher temperatures,  the linear-in-$T$ behavior is dominant before it
trends to vanish. It is in qualitative agreement with the direct measurement of superfluid density in films of Fe-pnictide superconductor at $x=0.08$. The evolution of $\Delta\lambda(T)$ can be roughly
fitted by a power-law function with the exponent depending on the
doping concentration. We show that the Uemura relation holds for the
iron-based superconductors only at very low doping levels.
\end{abstract}
 \maketitle

In addition to zero resistance, the Meissner effect is another hallmark of
superconductivity. The directly measured penetration depth($\lambda$) in a weak
magnetic field provides information of the gap structure, and is a
characteristic length scale of a bulk superconductor. In general, $\rho_s\propto1/\lambda^2$.
The number of electrons in the superconducting phase, $\rho_s$, characterizes the phase
rigidity of a superconductor. In conventional Bardeen-Cooper-Schrieffer (BCS) superconductors, the penetration depth exhibits
an exponential behavior at low temperatures, and the power-law
behavior in $\Delta\lambda(T)\equiv \lambda(T)-\lambda(0)$ has been considered as evidence for
unconventional pairing symmetry in the high-temperature
superconductors~\cite{4}.
Compare to cuprates, the remarkable features of
iron pnictides are the nature of magnetism and the
multiband character. They have triggered massive
studies since their
discovery~\cite{q1,q3}. In this letter we focus on its response to a weak external magnetic field.

There are several ways to measure magnetic
penetration depth~\cite{xu,exp2,mudiff}.
In the $1111$ systems, at low temperatures, some experiments~\cite{exp5}
found a power-law behavior $\lambda(T)$, while others~\cite{exp6,exp7} have found an exponential temperature dependence of $\lambda(T)$. The situation in the 122 system is also
unclear: The superfluid density $\rho_s(T)$ exhibits an exponential behavior in
the cleanest $\mathrm{Ba_{1-x}K_xFe_2As_2}$~\cite{Fegap2},  while
 measurements on $\mathrm{Ba(Fe_{1-x}Co_x)_2As_2}$ have
shown a power-law behavior of
$\lambda(T)$~\cite{Fegap3,exp4,exp9,jiey,aabar,lamd04} with the
exponent varying from $1.6$ to $2.8$, and a two-gap scenario is
suggested for $\mathrm{Ba(Fe_{1-x}Co_x)_2As_2}$ and
$\mathrm{Ba_{1-x}Rb_xFe_2As_2}$~\cite{ted3,exp1}.
And there are also some theoretical works~\cite{bakfe,rafael,yunkyu,ro1}.

In this letter, we carry out systematic calculations of $\rho_s(T)$ based on a two-orbital phenomenological
model~\cite{zhang}. Within this model, each unit cell accommodates two
inequivalent Fe ions and results based
on this model on various properties of Fe-pnictide superconductors~\cite{zhang,zhou,huang,gao1,gao2,huang2,huang3,gao3,zhou3} are in reasonable agreement with experimental measurements. When we normalize the energy parameters of the Fe-Fe nearest and next-nearest neighbors, the hopping integrals defined below are chosen as
$t_{1-4}=1, 0.4, -2.0, 0.04$~\cite{zhang}, respectively. In the momentum $k$ space, the single-particle Hamiltonian matrix can be written as~\cite{gao1,huang2}
 \begin{eqnarray}\label{ht}
H_{t,k}&=&\left(
  \begin{array}{cccc}
    a_1-\mu & a_3 & a_4 & \,0 \\
    a_3 & a_1-\mu & \,0 & a_4 \\
    a_4 & \,0 & a_2-\mu & a_3 \\
    \,0 & a_4 & a_3 & a_2-\mu \\
  \end{array}
\right),
\end{eqnarray}
 with $a_1=-2t_2\cos{(k_x+k_y)}-2t_3\cos{(k_x-k_y)}$,
$a_2=-2t_3\cos{(k_x-k_y)}-2t_2\cos{(k_x+k_y})$,
$a_3=-2t_4(\cos{(k_x+k_y)}+\cos{(k_x-k_y)})$,
$a_4=-2t_1(\cos{k_x}+\cos{k_y})$, where $\mu$ is the chemical potential.
Here we have chosen the $x$ axis along the link connecting nearest
neighbor (NN) Fe ions, and the distance between NN Fe is taken as the unit of length. The pairing term $H_{\Delta,k}=\sum_{\alpha\nu
\bf{k}}(\Delta_{\alpha,\bf{k}}c^{\dag}_{\alpha\nu
\bf{k}\uparrow}c^{\dag}_{\alpha\nu -\bf{k}\downarrow}+H.c.)$ has only
next-nearest-neighbor (NNN) intra-orbital pairing, where $\alpha$ denotes Fe $A$ or Fe $B$ in the unit cell and $\nu$ denotes the orbitals. It will lead to the
$s_{\pm}$-wave pairing symmetry~\cite{Fegap2,Fegap3,Fegap4}.
The self-consistent conditions are
$\Delta_{\alpha\bf{k}}=2\sum_{\tau}\cos{\bf{k}}_{\tau}\Delta^{\alpha}_{i,i+\tau}$
and $\Delta^{\alpha}_{i,i+\tau}=\frac{V}{2}\langle
c^{\alpha}_{i\nu\uparrow}c^{\alpha}_{i+\tau,\nu
\downarrow}-c^{\alpha}_{i\nu\downarrow}c^{\alpha}_{i+\tau,\nu
\uparrow} \rangle =\frac{V}{N_s}\sum_
{\bf{k}}\cos{\bf{k}}_{\tau}\langle c_{\alpha\nu,
\bf{k}\uparrow}c_{\alpha\nu, -\bf{k}\downarrow} \rangle$, with
$\tau=\bf{x}\pm \bf{y}$ and the pairing strength $V=1.2$. The
interaction term includes the Hund's coupling $J_{H}=1.3$ and the
on-site Coulomb interaction $U$, in which we choose $U=3.4$ and
$U=4.0$ as two different kinds of homogenous systems. After taking
the mean-field treatment~\cite{zhou,huang}, $H_{int}$ can be
expressed as
\begin{eqnarray}\label{2}
H_{int}&=&U\sum_{i\mu\sigma\neq\bar{\sigma}}\langle n_{{
i}\mu\bar{\sigma}}\rangle n_{{i}\mu\sigma}+(U-3J_H)\sum_{
i\mu\neq\nu\sigma} \langle n_{i\mu\sigma}\rangle n_{i\nu\sigma}\nonumber\\
&&+(U-2J_H)\sum_{i\mu\neq\nu\sigma\neq\bar{\sigma}} \langle n_{
i\mu\bar\sigma}\rangle n_{i\nu \sigma}.
\end{eqnarray}
In the presence of spin-density-wave ($\mathrm{SDW}$) order,  $H_{int}$ in the $k$ space can
be decoupled into a diagonal term and magnetic term. Define
$\psi^{\dag}_{\bf{k}\sigma}=(c^{\dag}_{A0,\bf{k}\uparrow},c^{\dag}_{A1,\bf{k}\uparrow},c^{\dag}_{B0,\bf{k}\uparrow},c^{\dag}_{B1,\bf{k}\uparrow})$,
$\varphi^{\dag}_{\bf{k}}=(\psi^{\dag}_{\bf{k}\uparrow},\psi^{\dag}_{\bf{k+Q}\uparrow},\psi_{-\bf{k}\downarrow},\psi_{-\bf{k+Q}\downarrow})$,
the Hamiltonian without external field in $k$ space can be written as
$\varphi^{\dag}_{\bf{k}} H_0
\varphi_{\bf{k}}$~\cite{gao1,huang2}, with
\begin{eqnarray}\label{3}
H_0&=&\left(\begin{array}{cccc}
 H^{\prime}_{t,\bf{\bf{k}}}    & R   &  IH_{\Delta,\bf{\bf{k}}}    & 0\\
 R  & H^{\prime}_{t,\bf{k+Q}}   &  0            & IH_{\Delta,\bf{k+Q}}\\
  IH_{\Delta,\bf{k}}    & 0              & -H^{\prime}_{t,\bf{k}}  &R\\
     0         & IH_{\Delta,\bf{\bf{k+Q}}}       &  R &-H^{\prime}_{t,\bf{\bf{k+Q}}}\\
\end{array}\right),
\end{eqnarray}
where $I$ is a $4\times 4$ unit matrix, $R=-\frac{M}{2}(U+J_H)H_M$,
and the corresponding
$H^{\prime}_{t,\bf{k}}=H_{t,\bf{k}}+\frac{n}{4}(3U-5J_H)I$, with
$n=2+x$. $R$ relates to the magnetic
order~\cite{gao1,huang2} with
\begin{eqnarray}\label{3}
 H_{M}&=&\left(\begin{array}{cc}
 \mathcal{I}& 0 \\
 0& \mathcal{I}\exp{\mathrm{i}\bf{Q}\cdot \textbf{R}_{AB}} \\
\end{array}\right)\;,
\end{eqnarray}
in Eq.(4) $\mathcal{I}$ is a $2\times 2$ unit matrix.
Due to $\mathrm{SDW}$ order, the wave vector $\bf{k}$ is restricted in the magnetic Brillouin zone (BZ).
The self-consistent condition is
$M=\frac{1}{2}\sum_{\nu}(n_{A\nu\uparrow}-n_{A\nu\downarrow})=\frac{1}{2N_s}\sum_{\nu,\bf{k}} \sigma c^{\dag}_{A\nu\sigma
\bf{k}}c_{A\nu\sigma \bf{\bf{k+Q}}}$, $\textbf{R}_{AB}$ is the distance of Fe B to the origin sited by Fe A. $N_s$ is the number of unit
cells. We take $N_s=512$ to obtain self-consistent
 parameters and $N_s=768$ in the calculation of $\rho_s$. After diagonalizing $\sum_{\bf{k}}\varphi^{\dag}_{\bf{k}} H_0
\varphi_{\bf{k}}=\sum_{\bf{k}m}E_{\bf{k},m}\gamma^{\dag \bf{k}}_{m}\gamma^{\bf{k}}_{m}$  by a
 $16\times 16$ canonical transformation matrix $\mathbb{T}$, we can obtain all properties of the system
 without the external field.

Our investigation of the superfluid density $\rho_s$ follows the
linear response approach described by Refs.~\cite{1,2,3,4}.
In the presence of a slowly varying vector potential
$A_x(r,t)=A(q,\omega)e^{\mathrm{i}\bf{q}\cdot r_i-\mathrm{i}\omega
t}$ along the $x$ direction, the hopping term is modified by a phase
factor, $c^{\dag}_{i\sigma}c_{j\sigma}\rightarrow
c^{\dag}_{i\sigma}c_{j\sigma}\exp{\mathrm{i}\frac{e}{\hbar
c}\int^{r_i}_{r_j}\textbf{A}(\textbf{r},t)\cdot \mathrm{d} \textbf{r}} $. Throughout the letter we set
$\hbar=c=1$. By expanding the factors to the order of $A^2$, we
obtained  the total Hamiltonian $H_{tot}=H_0+H^{\prime}$ with
\begin{eqnarray}\label{5}
H^{\prime}=-\sum_iA_x(r_i,t)[eJ^P_x(r_i)+\frac{1}{2}e^2A_x(r_i,t)K_x(r_i)].
\end{eqnarray}
$J^P_x(r_i)$ is the particle current density along the $x$ axis, $K_x(r_i)$ is
the kinetic energy density along the $x$ axis. Their expressions are
\begin{eqnarray}\label{5.1}
K_x(r_i)&=&-\sum_{\nu\nu^{\prime}\sigma\delta}t_{i,i+\delta}x^2_{i,i+\delta}(c^{\dag}_{i\nu\sigma}c_{i+\delta,\nu^{\prime}\sigma}+H.c.),\\
J^P_x(r_i)&=&-\mathrm{i}\sum_{\nu\nu^{\prime}\sigma\delta}t_{i,i+\delta}x_{i,i+\delta}(c^{\dag}_{i\nu\sigma}c_{i+\delta,\nu^{\prime}\sigma}-H.c.),
\end{eqnarray}
only $\delta=x,x\pm y$ have contributions to the $x$ component and
$x_{i,i+\delta}=1$ in our coordination. The charge current density along the $x$ axis is defined as
\begin{eqnarray}\label{6}
J^{Q}_x(r_i)\equiv-\frac{\delta H^{\prime}}{\delta
A_x(r_i,t)}=eJ^{p}_x(r_i)+e^2K_x(r_i)A_x(r_i,t).
\end{eqnarray}
The kinetic energy is calculated to zeroth order of $A_x(r_i)$,
corresponding to the diamagnetic part, and that of the paramagnetic
part $J^{P}_{x}(r_i)$ is calculated to the first order of
$A_x(r_i)$. In the interaction representation we have
\begin{eqnarray}\label{10}
\langle J^{P}_{x}(r_i)\rangle &=&-\mathrm{i}\int_{-\infty}^{t}
\langle [J^{P}_{x}(r_i,t),H^{\prime}(t^{\prime})]_{-} \rangle_0dt^{\prime}\nonumber\\&=&-\frac{eA_x(r,t)}{N_s}\Pi_{xx}(\bf{q},\omega),
\end{eqnarray}
$\langle\rangle$ represents the expectation value based on the wave
function of $H_{tot}$ while $\langle\rangle_0$ corresponds to the
wave function of $H_0$. In the Matsubara formalism we have the
current-current correlation $\Pi_{xx}(\textbf{q},\mathrm{i}\omega)=
\int^{\beta}_0 d\tau
  e^{\mathrm{i}\omega\tau}\Pi_{xx}({\bf{q}},\tau)$, and $\Pi_{xx}(\textbf{q},\tau)=-\langle T_{\tau} J^P_x(\textbf{q},\tau)J^P_x(-\textbf{q},0)
\rangle_0=\sum_{m_1m_2}\Pi^{m_1m_2}_{xx}(\textbf{q},\tau)$ where $\mathrm{T}_{\tau}$ is the time ordering operator,
$J^P_x(\textbf{q},\tau)=e^{\tau H_0}J^P_x(\textbf{q}) e^{-\tau H_0}$,
$J^P_x(\textbf{q})=\sum_{i}e^{-\mathrm{i}\textbf{q}\cdot
\textbf{r}_i}J^P_x(r_i)=\sum_{m_1m_2}J^P_{m_1,m_2}(\textbf{q})$ is a summation over $\textbf{k}$. Calculation of
$\Pi_{xx}(\textbf{q},\mathrm{i}\omega)$
is in the framework of equations of motion of Green's function,
\begin{eqnarray}
\frac{d\Pi^{m_1m_2}_{xx}(\textbf{q},\tau)}{d \tau}&=&-[J^P_{m_1,m_2}(\textbf{q}),J^P_{x}(-\textbf{q})]_{-}\nonumber\\
&-&\langle T_{\tau}e^{H_0\tau}[H_0,J^P_{m_1,m_2}(\textbf{q})]_{-}e^{-H_0 \tau} J^P_{x}(-\textbf{q},0) \rangle_0 .\nonumber
\end{eqnarray}
A lengthy but straightforward algebra leads to
\begin{eqnarray}\label{corr}
\Pi_{xx}(\textbf{q},\mathrm{i}\omega)
            \!\!=\!\!\!\!\!\sum_{{\bf{k}}m_1m_2} \!\! \frac{ Y_{m_1m_2}^{\bf{k},\bf{k+q}}Y_{m_2m_1}^{\bf{k+q},\bf{k}}(f(E_{{\bf{k}},m_1})-f(E_{{\bf{k+q}},m_2}) ) }{\mathrm{i}\omega+(E_{{\bf{k}},m_1}-E_{{\bf{k+q}},m_2})},
\end{eqnarray}
where $f$ is the Fermi distribution function. Through analytic continuation,
 $\Pi_{xx}(\textbf{q},\omega)$ is obtained. When $\omega=0$, the
derivative of $f$ has an important contribution to
$\Pi_{xx}(q,\mathrm{i}\omega)$.
 The quantity $Y^{\textbf{k},\textbf{k+q}}_{m_1m_2}$ can be expressed as
 \begin{eqnarray}
Y^{\textbf{k},\textbf{k+q}}_{m_1m_2}&=&\frac{2}{N_s}[t_4(\xi_4(\sin{k_{x-y}}+\sin{k_{x+y}})+\xi^{\prime}_4(\sin{k^{\textbf{Q}}_{x-y}}+\sin{k^{\textbf{Q}}_{x+y}}) )\nonumber\\
&+&t_3(\xi_2\sin{k_{x-y}}+\tilde{\xi}_2\sin{k_{x+y}}+\xi^{\prime}_2\sin{k^{\textbf{Q}}_{x-y}}+\tilde{\xi}^{\prime}_2\sin{k^{\textbf{Q}}_{x+y}})\nonumber\\
&+&t_2(\xi_2\sin{k_{x+y}}+\tilde{\xi}_2\sin{k_{x-y}}+\xi^{\prime}_2\sin{k^{\textbf{Q}}_{x+y}}+\tilde{\xi}^{\prime}_2\sin{k^{\textbf{Q}}_{x-y}})\nonumber\\
&+&t_1(\xi_1\sin{k_x}+\xi^{\prime}_1\sin{k^{\textbf{Q}}_x})],
\end{eqnarray}
 with
$\xi_1=\alpha^{\textbf{k},\textbf{k+q}}_{1,3}+\alpha^{\textbf{k+q},\textbf{k}}_{3,1}
        +\alpha^{\textbf{k+q},\textbf{k}}_{9,11}+\alpha^{\textbf{k},\textbf{k+q}}_{11,9}$,
$\xi_2=\alpha^{\textbf{k},\textbf{k+q}}_{1,1}+\alpha^{\textbf{k},\textbf{k+q}}_{9,9}$,
$\tilde{\xi}_2=\alpha^{\textbf{k},\textbf{k+q}}_{3,3}+\alpha^{\textbf{k},\textbf{k+q}}_{11,11}$,
$\xi_4=\alpha^{\textbf{k},\textbf{k+q}}_{1,2}+\alpha^{\textbf{k+q},\textbf{k}}_{2,1}+\alpha^{\textbf{k+q},\textbf{k}}_{9,10}+\alpha^{\textbf{k},\textbf{k+q}}_{10,9}$,
and
$\alpha^{\textbf{k},\textbf{k}^{\prime}}_{ij}=\mathbb{T}^*_{i,m_1}(\textbf{k})\mathbb{T}_{j,m_2}(\textbf{k}^{\prime})+\mathbb{T}^*_{i+1,m_1}(\textbf{k})\mathbb{T}_{j+1,m_2}(\textbf{k}^{\prime})$.
The corresponding $\xi^{\prime}_i$ is connected to $\xi_i$  by changing
$\alpha_{i,j}$ into $\alpha_{i+4,j+4}$. $k_{x\pm y}$ denotes $k_x\pm k_y$ and $k^{\textbf{Q}}_{x\pm y}=k_{x\pm y}+\textbf{Q}$. The superfluid weight
measures the ratio of the superfluid density to the mass $D_s/\pi
e^2=\rho_s/m^{\ast}=-\langle J^{Q}_{x}(r_i,t) \rangle/e^2A_x(r_i)$,
and the Drude weight is a measurement of the ratio of density of mobile
charges to their mass~\cite{1,2,3,4},
\begin{eqnarray}\label{13}
\frac{D_s}{\pi e^2}&=&\frac{1}{N}\Pi_{xx}(q_x=0,q_y\rightarrow 0,\omega=0)-\langle K_x \rangle_0,\\
\frac{D}{\pi e^2}&=&\frac{1}{N}\Pi_{xx}(q_x=0,q_y=0,\omega\rightarrow 0)-\langle K_x \rangle_0.
\end{eqnarray}

\begin{figure}
\centering
    \vspace{-0.5cm}
      \includegraphics[width=9cm]{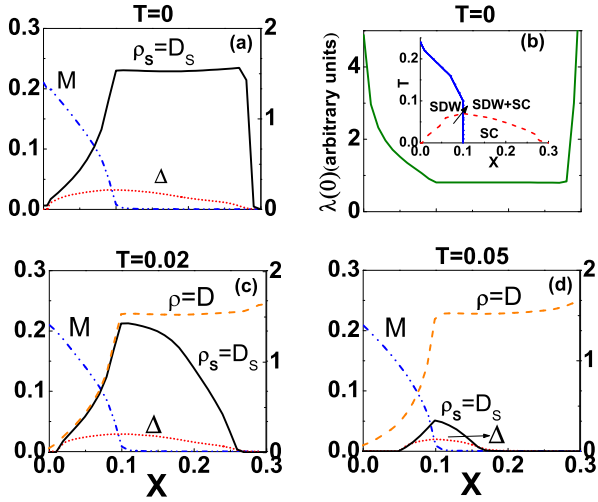}
      \vspace{-0.5cm}
\caption{(color online) Panels (a), (c), and (d) plot $D_s$ (black
solid line), $D$ (orange dashed line ), $\Delta$ (red dotted line),
and $M$ (blue dash-dot-dotted line) as functions of $x$ at different
temperatures. The right scale is for $D_s$ and $D$ while the left
scale is for $\Delta$ and $M$. Panel (b) plots $\lambda(0)$ as a
function of $x$. The inset of panel (b) is the phase diagram of
temperature $T$ and $x$. }\label{figDs}
\end{figure}

Figure~\ref {figDs} shows the variation of $D_s$, $D$,
$M$ and superconducting (SC) order
$\Delta=\frac{1}{4}\sum_{\alpha}(\Delta^{\alpha}_{i,i+{\bf{x+y}}}+\Delta^s_{i,i+{\bf{x-y}}})$,
as functions of $x$ at different temperatures. $D$ does not change much as the
temperature varies and we plot it clearly in Figs.~\ref {figDs}(c)
and ~\ref {figDs}(d). At zero temperature, we do not show the plot of $D$  because in almost all the doping
levels $D_s=D$ as long as $\Delta$ has finite value; Fig.~\ref {figDs}(a) shows that in the
overdoped regime, the superconducting gap disappears and $D_s$ drops to zero, while $D$ is finite just like the plot in panels (c) and (d);
hence, in the overdoped levels when $\Delta=0$ the system corresponds to metal.
We can see from Fig.~\ref {figDs}(a)
that at $T=0$, $D_s$ increases with the increase of $x$ until it reaches the SDW boundary. In the underdoped region $x<0.05$, most of the Fermi surfaces are gapped by SDW~\cite{zhou,huang3}, doping is the major source of charge carrier; hence, the superfluid density as well as mobile charge density increase linearly with the increase of $x$. While at larger doping $0.5<x<0.1$, SDW is suppressed, the gapped surfaces shrinks significantly, and more intrinsic charge carriers are released to the system in addition to the doping carriers. This is the reason why the increase of $D_S=D$ with doping becomes more dramatic than the linear dependence in this region.
After
SDW disappears, $\Delta$ dominates the behavior of $D_s$, and shows a flat behavior in a
considerably large doping range.
In panel (b) we show the variation
of $\lambda(0)$ as a function of $x$ for $x\leq0.3$. We define $\rho_s(T)=D_s(T)=\lambda(T)^{-2}$ with arbitrary units. Compared to the phase diagram
in the inset, we find that in the $\mathrm{SDW+SC}$ coexisting regime, $\lambda(0)$ shows a sharp increase with the
decrease of $x$, which is in
good agreement with experiments~\cite{exp4,exp9}.

An external magnetic field can couple relevant correlation
functions; hence, $\rho_s$ is a nonlocal quantity,
describing the stiffness of the system. Figure~\ref{figDs}(c) and ~\ref{figDs}(d) show that at finite $T$, $D_s$ deviates from $D$, the suppression of $D_s$ is stronger than that
of $\Delta$. For the
$U=4$ case, the results (not shown here) are very similar to the results presented here.

\begin{figure}
\centering
      \includegraphics[width=3.0cm]{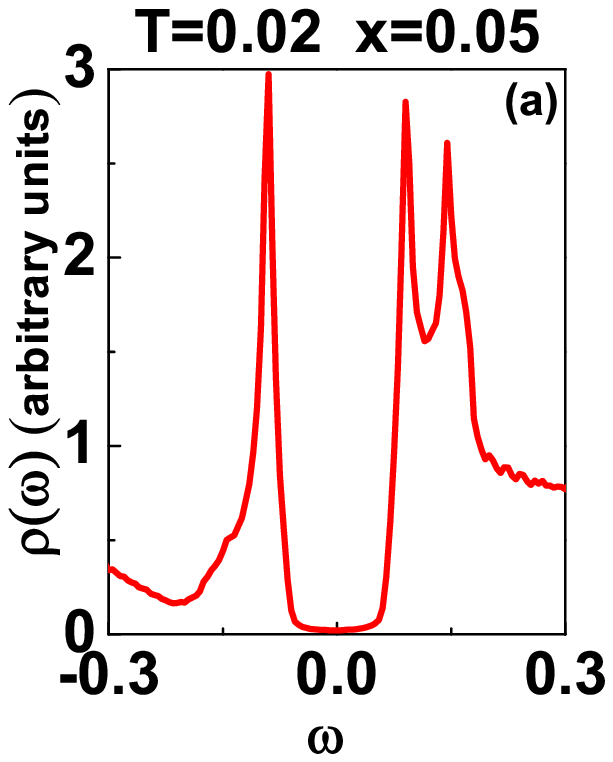}
      \includegraphics[width=2.7cm]{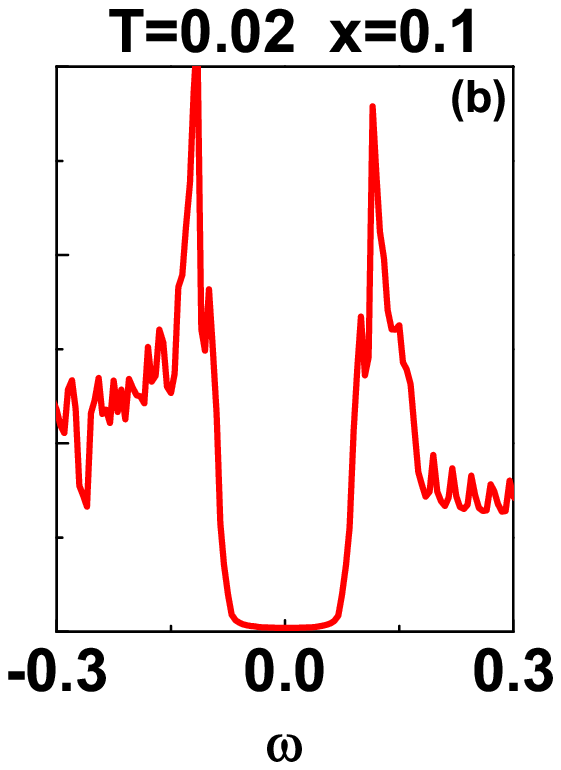}
      \includegraphics[width=2.7cm]{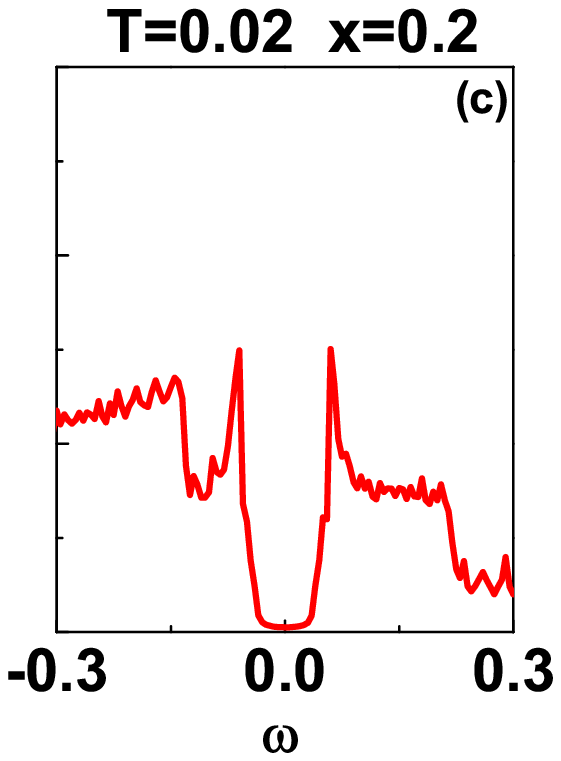}
\caption{(color online) Density of states at $T=0.02$ for different $x$. All those
calculations are for the $U=3.4$ case. }\label{figlos}
\end{figure}

Temperature dependence of superfluid density is a quantity
reflecting the low-energy residual density of states(DOS) inside the superconducting gap. Equation(\ref{corr}) indicates that the difference
between $D$ and $D_s$ is related to the derivation of $f$ near the Fermi surface, and can be understood as excitation of quasiparticles $\rho_q$.
Fig.~\ref{figlos} shows the DOS at
$T=0.02$. For $x=0.05$ and $0.1$ the gap is considerably larger, hence
$D_s$ is equal or almost equal to $D$. Although there is a gap at $x=0.2$[see Fig.~\ref{figlos}(c)], it is small; therefore, $f'(E_k)$
has its contribution to $D_s$, and therefore $D_s$ deviates from $D$.

\begin{figure}
\centering
   \vspace{-0.5cm}
      \includegraphics[width=9.5cm]{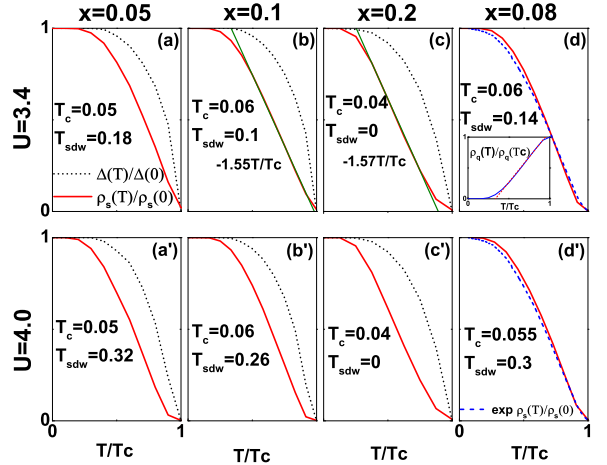}
  \vspace{-1cm}
\caption{(color online) Panels (a), (b), and (c) plot the
renormalized superfluid density $\rho_s(T)/\rho_s(0)$ and
superconducting order parameter $\Delta(T)/\Delta(0)$ as functions
of the temperature $T/T_c$ at different doping levels for $U=3.4$. $T_{sdw}$ is the transition temperature for SDW. The green dotted lines are linear-in-T fitting functions.
Panels ($a^{\prime}$), ($b^{\prime}$), and ($c^{\prime}$) are similar but for $U=4.0$. Panel (d),($d^{\prime}$) show the comparison of our results with experiment data at $x=0.08$.
Blue solid line in the inset of panel ($d^{\prime}$) plots $\rho_q(T)/\rho(T_c)$ as a function of $T/T_c$ at $x=0.08$ and the red dashed line is the aid for the eyes. }\label{figdepth}
\end{figure}

We choose three typical doping levels, to show the temperature $T/T_c$
dependence of $\rho_s(T)/\rho_s(0)$ and $\Delta(T)/\Delta(0)$ for $U=3.4$ as well as for $U=4.0$.
From Fig.~\ref{figdepth} we can see that the suppression of superfluid
density is stronger than that of the superconducting order parameter
in all cases. At low temperatures, the curve of
$\rho_s(T)/\rho_s(0)$ is flat, a characteristic of a nodeless
superconducting gap.

As $T$ increases, a linear-in-$T$ behavior
of superfluid density is dominant in all cases. For $U=3.4$ cases, linear functions $-1.55T/T_c+1.52$ and $-1.57T/T_c$+1.49 are used to
fit this kind of behavior for $x=0.1$ and $x=0.2$, respectively, which are shown in Figs.~\ref{figdepth}(b) and ~\ref{figdepth}(c). It is
consistent with the power-law behavior observed in the
experiments~\cite{Fegap3,exp4,exp9,jiey,aabar,lamd04}.
Interestingly, they are in good agreement with the direct measurements of superfluid density in films of Fe-pnictide superconductors in Ref.\onlinecite{jiey}.
We show our results and the experimental data [see Fig.1(a) in Ref.\onlinecite{jiey}] together in Figs.\ref{figdepth}(d) (U=3.4 case) and 3($d^{\prime}$) (U=4.0 case), and their consistence is explicit. In order to understand the wider linear $T$ dependence of $\rho_s(T)$, the
inset in Fig.~\ref{figdepth}(d) plots the renormalized $\rho_q(T)/\rho(T_c)$ as a function of $T/T_c$ at $x=0.08$;
the red dashed line aids for eyes. We can see that the number of excited quasiparticles is exponentially small at low $T$ with strong superconductivity, but it is proportional to linear $T$ within a certain temperature range before superconductivity disappears. The easy appearance of
 linear-in-T behavior is closely related to anisotropic $S_{\pm}$ superconducting paring, since in-gap states(Andreev states) may be induced in this case.
 The ratio $2\Delta_k(0)/k_BT_c$ at optimal doping is about $4.3\; (4.5)$ for the
$U=3.4\;(4.0)$ system.

\begin{figure}
\centering
      \includegraphics[width=4.1cm]{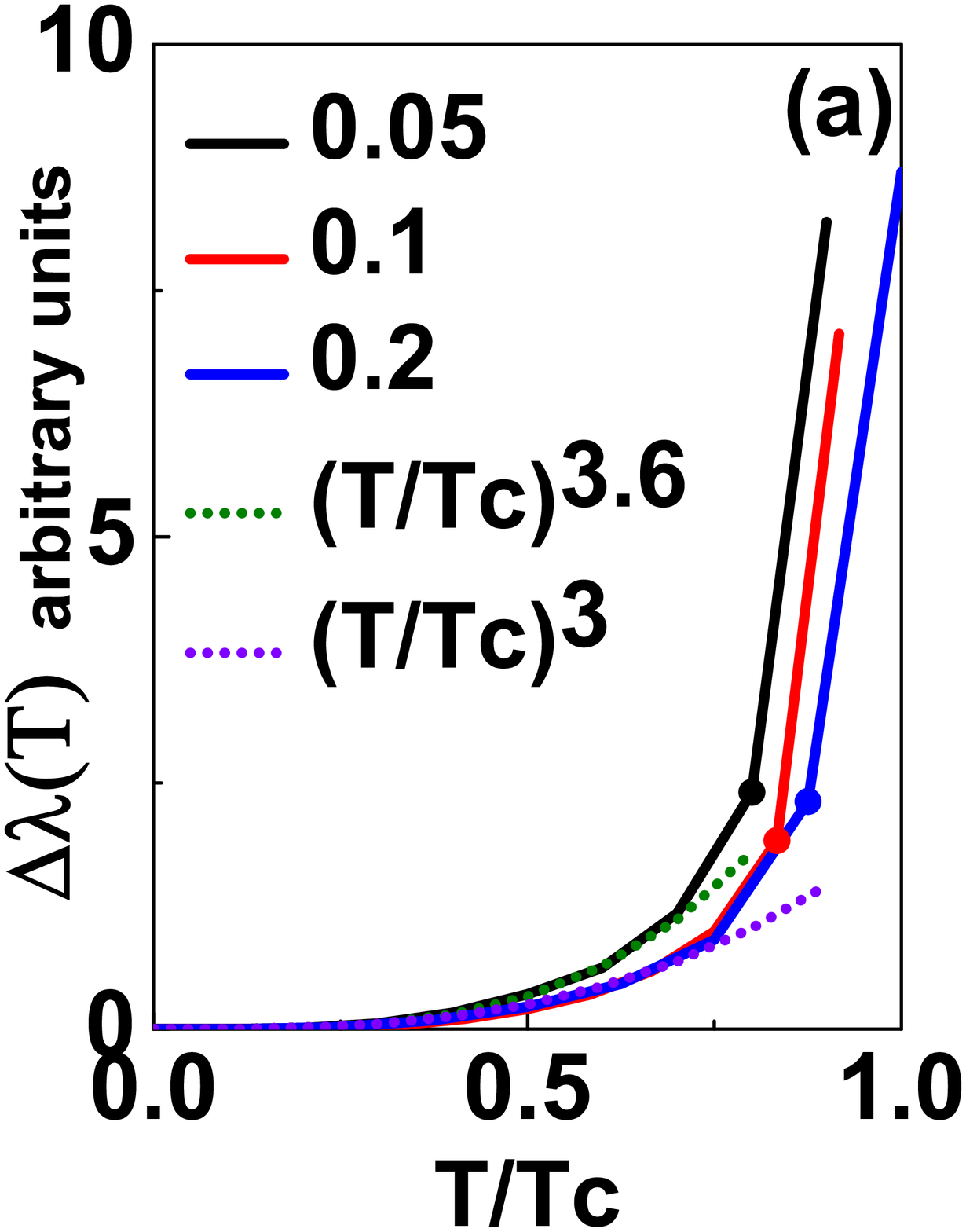}
      \includegraphics[width=4.1cm]{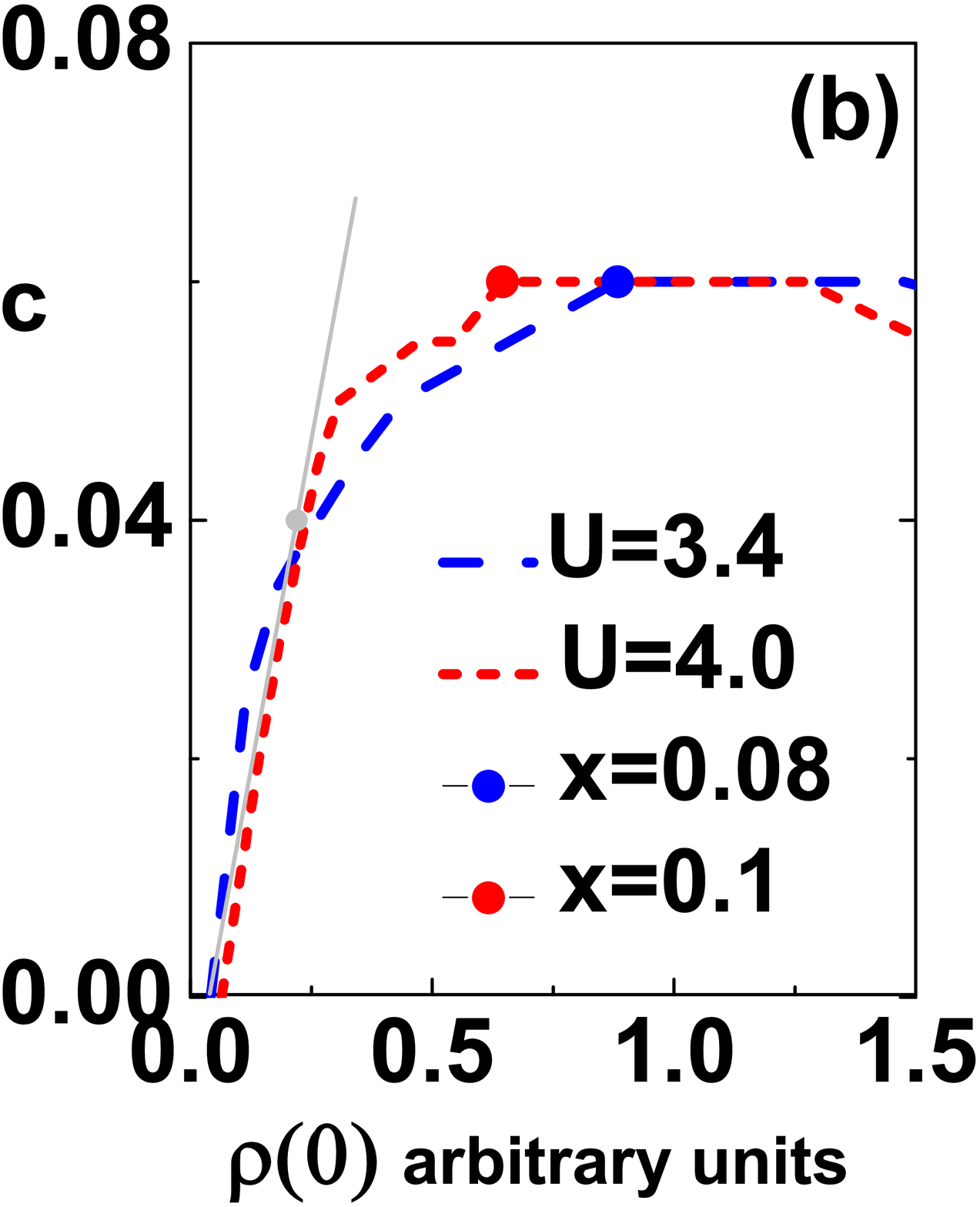}
\caption{(color online) Panel (a) plots $\Delta \lambda(T)$ as a function of $T/T_c$ at
typical selected doping for $U=4$, the dashed lines are the corresponding fitting functions.
Panel (b) is the Uemura plot of Fe-base superconductor. The
$x$ axis is $\rho_s(0)$ for different doping, the $y$ axis is the corresponding
$T_c$ for the given dopings.}\label{figuemu}
\end{figure}

Experiments always measure $\Delta\lambda(T)=\lambda(T)-\lambda(0)$,
so we show the evolution of  $\Delta\lambda(T)$ at selected doping
concentrations for $U=4.0$ in Fig.~\ref{figuemu}(a). The results of
$U=3.4$ are very similar. In the low-temperature range the curve is
flat. At high
temperature approaching the disappearance of superconductivity,
there is a jump for the value of $\Delta\lambda(T)$, which we show
by the colored solid dots. We fit the evolution of
$\Delta\lambda(T)$ by a power-law behavior. See
Fig.~\ref{figuemu}(a); the corresponding fitting function $4(T/T_c)^{3.6}$($2(T/T_c)^{3}$ ) is for data of $x=0.05$ ($x=0.1,0.2$) and it may be the reason why the experiments give different exponents for different samples.

Experiments have shown that the Uemura relation~\cite{Uerelation} holds
~\cite{uehold} for a 1111 system but does not hold for a 122 system ~\cite{ueunhold}. In
Fig.~\ref{figuemu}(b), we plot $T_c$ versus $\rho_s(0)$ based on our
model. The blue-dashed line (red-dotted line) is for the $U=3.4$
($U=4.0$) system. It shows that at very low doping levels, about $x<0.035$(grey point), both the
$U=3.4$ and $U=4$ systems follow the same empirical linear relation(grey line).
As $T_c$ close to
the maximum and $\rho_s(0)$ saturate at $x>0.08$ ($0.1$) for $U=3.4$ ($U=4.0$), and the data significantly deviate from the linear
relation. This is because in the very underdoped region the doping is a major source of charge carriers and the Uemura relation is valid here.

Based on a two-orbital phenomenological model, we have studied the stiffness of superconductivity in
clean iron-based superconductors.
At zero temperature, we find $\lambda(0)$ a sharp jump as $x$ decreases in the regime of the coexisting $\mathrm{SDW+SC}$ orders;
 the variation of $\lambda(0)$ as a function of doping is in good agreement with experiments~\cite{exp4}. As far as we know this is a new theoretical result. At low temperatures, $\rho_s(T)/\rho_s(0)$ is flat, then shows a linear-in-T behavior before the system loses its superconductivity. It is in good agreement with experiments of direct measurement of superfluid density in films \cite{jiey}. The evolution of
 $\Delta\lambda(T)$ roughly follows the power-law behavior with different exponents corresponding to different doping levels. Only at low doping levels, the
empirical Uemura linear relation holds for the iron-based
superconductors.

This work was supported by the Texas Center for Superconductivity
at the University of Houston and by the Robert
Welch Foundation under Grant No. E-1146 (H.H, Y.G. , C.S.T.), and by
the NNSA of the U.S. DOE at LANL under Contract No. DE-AC52-06NA25396
and the U.S. Department of Energy Office of Basic Energy Sciences (J.-X.Z.), and by NSFC No.11204138(Y.G.).

\end{document}